\documentclass{appolb}
\usepackage{graphicx}
\usepackage{subfigure,amsmath}

\begin{document}
\title{Susceptibilities in the PNJL model with 8q interactions and comparison with lQCD%
\thanks{Presented at Internationational Symposium on Multiparticle Dynamics}%
}
\author{Jo\~{a}o Moreira, Brigitte Hiller, Alexander Osipov, Alex Blin
\address{Centre for Computational Physics, Department of Physics, University of Coimbra, 3004-516 Coimbra, Portugal}
\\
}
\maketitle
\begin{abstract}
We present some results pertaining quantities which are regarded as good indicators of the pseudo-critical temperatures for the deconfinement and partial chiral restoration transitions using a polynomial form for the Polyakov potential part and an extended version of the NJL model which includes 6 and 8 quark interaction terms. Some comparisons with results from the lattice formulation of QCD (lQCD) are performed and results for the location of the critical endoint in the phase diagram are also presented. It is shown that the comparison with lQCD results favors a moderately strong OZI-violating 8 quark interaction term.
\end{abstract}
 
\section{Introduction}

Low energy effective models are a valuable tool for the study of the low-energy thermodynamic properties of strongly interacting matter. Their importance is even greater in the case for finite chemical potential where the sign problem afflicts the more fundamental approach of using lQCD. 

Two of the main features of the non-perturbative regime of QCD are the spontaneous breaking of chiral symmetry and confinement. For both a transition is expected to occur at finite temperature and/or chemical potential: partial restoration of chiral symmetry and deconfinement (with the spontaneous breaking of the $Z(N_c)$ centre symmetry).

The NJL model \cite{Nambu:1961} incorporates by construction a mechanism for the dynamical breaking of chiral symmetry and shares with QCD its global symmetries. The unwanted axial symmetry can be removed by the inclusion of a 't Hooft determinantal \cite{Hooft:1976,Bernard:1988,Reinhardt:1988} term but this however introduces a vacuum stability problem when considering the light quark sector ($u$, $d$ and $s$ quarks). This issue can be solved by further extending the model to include eight-quark interaction terms \cite{Osipov:2006a,Osipov:2006b}.

A partial inclusion of the gluonic degrees of freedom can be done through the consideration of a static background homogeneous temporal gauge gluonic field coupled to the quark fields through the covariant derivative \cite{Fukushima:2004}. The Polyakov loop can then be used as an approximate order parameter for deconfinement (it vanishes for the confined phase). An additional temperature-dependent pure ``gluonic'' term, the Polyakov potential, $\mathcal{U}$, for which several forms have been proposed, must also be introduced to drive the transition. Due to space constraints here we will only present results referring to the polynomial form as proposed in \cite{Ratti:2006} (furthermore the results are qualitatively similar with the other forms tested).


The specification of an non-renormalizable effective model is not complete without the choice of a regularization procedure. In the present study we will use a Pauli-Villars regulator with two subtractions in the integrand \cite{Osipov:1985}. In \cite{Moreira:2011} we have shown that using this we can obtain the expected high-temperature asymptotic behaviour for the PNJL model while consistently using the same regularization procedure for both the vacuum and medium contributions for the relevant integrals .

The thermodynamic potential is given by (for more details see \cite{ Moreira:2011, Hiller:2010}):
\begin{align}
\label{Omega}
&{\Omega\left(M_f,T,\mu,\phi,\overline{\phi}\right)}=\nonumber\\
=& \frac{1}{16}\left.\left(4Gh_f^2+\kappa h_uh_dh_s+\frac{3g_1}{2}\left(h_f^2\right)^2+3g_2h_f^4\right)\right|_0^{M_f} \nonumber \\
+&\frac{N_c}{8\pi^2}\!\sum_{f=u,d,s}\!\!\left( {J_{-1}(M_f^2,T,\mu,\phi,\overline{\phi} )} + C(T,\mu )\right)+ {\mathcal{U}\left(\phi,\overline{\phi},T\right)}.
\end{align}
Here $G$, $\kappa$, $g_1$, $g_2$ refer to the coupling strengths of the 4q (NJL term), 6q ('t Hooft determinant) and OZI-violating and non-violating 8q interactions respectively. $M_f$ and $h_f$ ($f=u,d,s$) refer to the dynamical masses and the condensates which must satisfy the conditions ($m_f$ refers to current masses):
\begin{align}
\left\{
\begin{array}{l}
m_u-M_u=G h_u +\frac{\kappa}{16}h_d h_s +\frac{g_1}{4}h_u h_f^2+\frac{g_2}{2}h_u^3\\
m_d-M_d=G h_d +\frac{\kappa}{16}h_u h_s +\frac{g_1}{4}h_d h_f^2+\frac{g_2}{2}h_d^3\\
m_s-M_s=G h_s +\frac{\kappa}{16}h_u h_d +\frac{g_1}{4}h_s h_f^2+\frac{g_2}{2}h_s^3.
\end{array}
\right.
\end{align}
$J_{-1}$ and $C$ (a mass independent, but $T,\mu$ dependent, term which must be added to ensure thermodynamic consistency \cite{Moreira:2011,Hiller:2010}) are given by:
\begin{align}
J_{-1}\left(M^2,T,\mu,\phi,\overline{\phi}\right)&=J^{vac}_{-1}\left(M^2\right)+J^{med}_{-1}\left(M^2,T,\mu,\phi,\overline{\phi}\right)\nonumber\\
J^{vac}_{-1}\left(M^2\right)&=-\int^\infty_0\mathrm{d}p\,\hat{\rho}_\Lambda 8 p^2(E_M-E_0)\nonumber\\
J^{med}_{-1}\left(M^2,T,\mu,\phi,\overline{\phi}\right)&=\int^\infty_0\mathrm{d}p\,8p^2\hat{\rho}_\Lambda\left(2\left(E_M-E_0\right)+T\ln\frac{n_{q\,M}n_{\overline{q}\,M}}{n_{q\,0}n_{\overline{q}\,0}}\right)\nonumber\\
C(T,\mu)&=-\frac{8}{3}\int^{\infty}_0\mathrm{d}p p^4 \left(\frac{n_{q~0}+n_{\overline{q}~0}}{p}\right).
\end{align}
Here $E$ is the energy and $n_q$/$n_{\overline{q}}$ refer to the occupation numbers (the subscript $_0$ refers to the occupation number for massless particles). The extension to include the Polyakov loop in the $J_{-1}$ integrals can be done straightforwardly by noticing that its phase enters the action as an imaginary chemical potential and modifying the occupation numbers accordingly. The regulator was chosen to be $\hat{\rho}_\Lambda=1-\left(1-\Lambda^2 \partial_{M^2}\right)\exp\left(\Lambda^2\partial_{M^2}\right)$.

\section{Transition signals}

Several quantities are expected to undergo a rapid change during the transitions acting therefore as a signal for the determination of the pseudo critical temperatures (note that with physical quark current masses both the chiral and the $Z(N_c)$ centre symmetries are only approximate; they are only exact in the chiral and quenched limit respectively). Besides a rapid change in the condensates and Polyakov loop one also expects a distinct signal coming from the chiral, quark number and Polyakov loop susceptibilities:
\begin{align}
    \chi^{i}_{chi}&
		=-\frac{1}{T^2}\left(\left.\frac{\partial^2 \Omega}{\partial^2 m_i}\right|_T-\left.\frac{\partial^2 \Omega}{\partial^2 m_i}\right|_{T=0}\right),&
		\chi^{i}_{num}&=-\frac{1}{T^2}\frac{\partial^2\Omega}{\partial \mu_i^2},\nonumber\\
		\chi_{\phi}&=\frac{1}{4}\left(\frac{\partial^2\Omega^\prime}{\partial\eta^2}+2\frac{\partial^2\Omega^\prime}{\partial\eta\partial\overline{\eta}}+\frac{\partial^2\Omega^\prime}{\partial\overline{\eta}^2}\right).&
  \end{align}

The use of the dual chiral condensates has been suggested as a possible additional indicator for deconfinement \cite{Bilgici:2008}. These correspond to the Fourier transform with respect to an angle which is introduced through the consideration of twisted temporal boundary conditions: 
\begin{align}
\Sigma^{(j)}_{i}=\int^{+\pi}_{-\pi} \frac{e^{-\imath \alpha j}}{2\pi} h_i(\alpha)\mathrm{d} \alpha,
\end{align}
where $h_i(\alpha)$ is given by $\frac{\partial\Omega}{\partial h_i}=0$ with $\mu\rightarrow\mu+\imath T\alpha$, (with $-\pi\leq \alpha<\pi$). Results for the temperature dependence of the dressed Polyakov loop (which corresponds to setting $j=1$ in the above) will also be presented here.

\section{Results}

As we reported in \cite{Osipov:2006b} it is possible to obtain the same low energy scalar and pseudoscalar mesonic spectra throughout a large range of values for the OZI-violating 8q coupling strength by simultaneously lowering G when increasing $g_1$ (except for the $\sigma$ meson mass which decreases for higher $g_1$). Here we will focus therefore in the impact of this choice in the transitions behaviour. Another parameter, $T_0$, coming from the polynomial form of $\mathcal{U}$ has also to be taken into account (it sets the temperature scale for the temperature dependence of $\mathcal{U}$).

As can be seen in Fig. 1 the chiral condensates, the Polyakov loop and the dual chiral condensates undergo a rapid change in a narrow temperature range signaling the transitions. The change is more rapid and occurs at a lower temperature for stronger 8q interactions. The behavior of the dressed Polyakov loop closely follows that of the chiral condensate and the zeros of the second derivative with respect to the temperature are in fact very close to each other (a few $\mathrm{MeV}$ apart at most). The temperature dependence of the susceptibilities is also strongly influenced by the value of $g_1$ as can be seen in Fig. 2: the peaks get sharper and higher with increasing $g_1$
\begin{figure}[htp]
  \centering
  \subfigure[]{\includegraphics[width= 0.3 \textwidth]{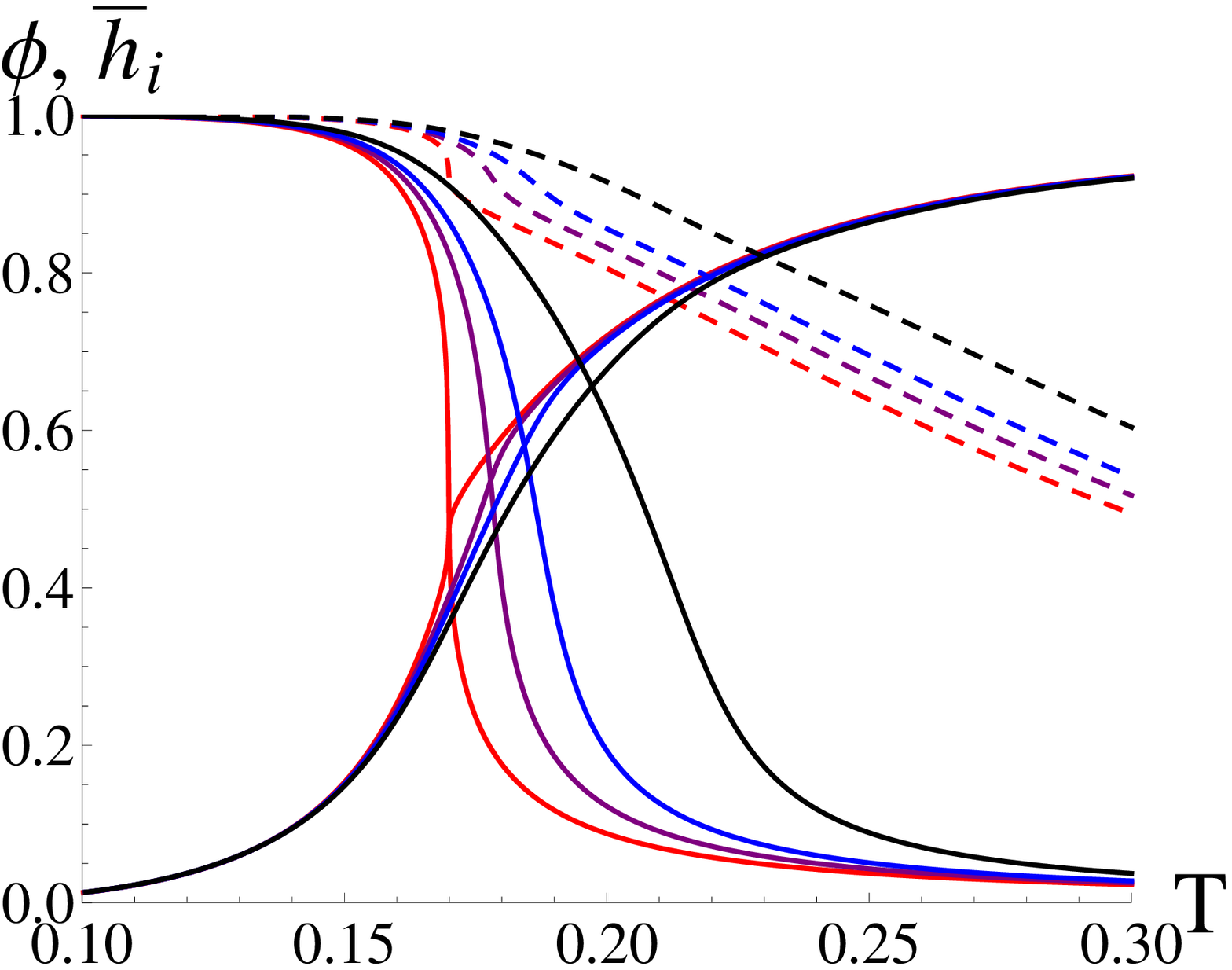}}
  \subfigure[]{\includegraphics[width= 0.3 \textwidth]{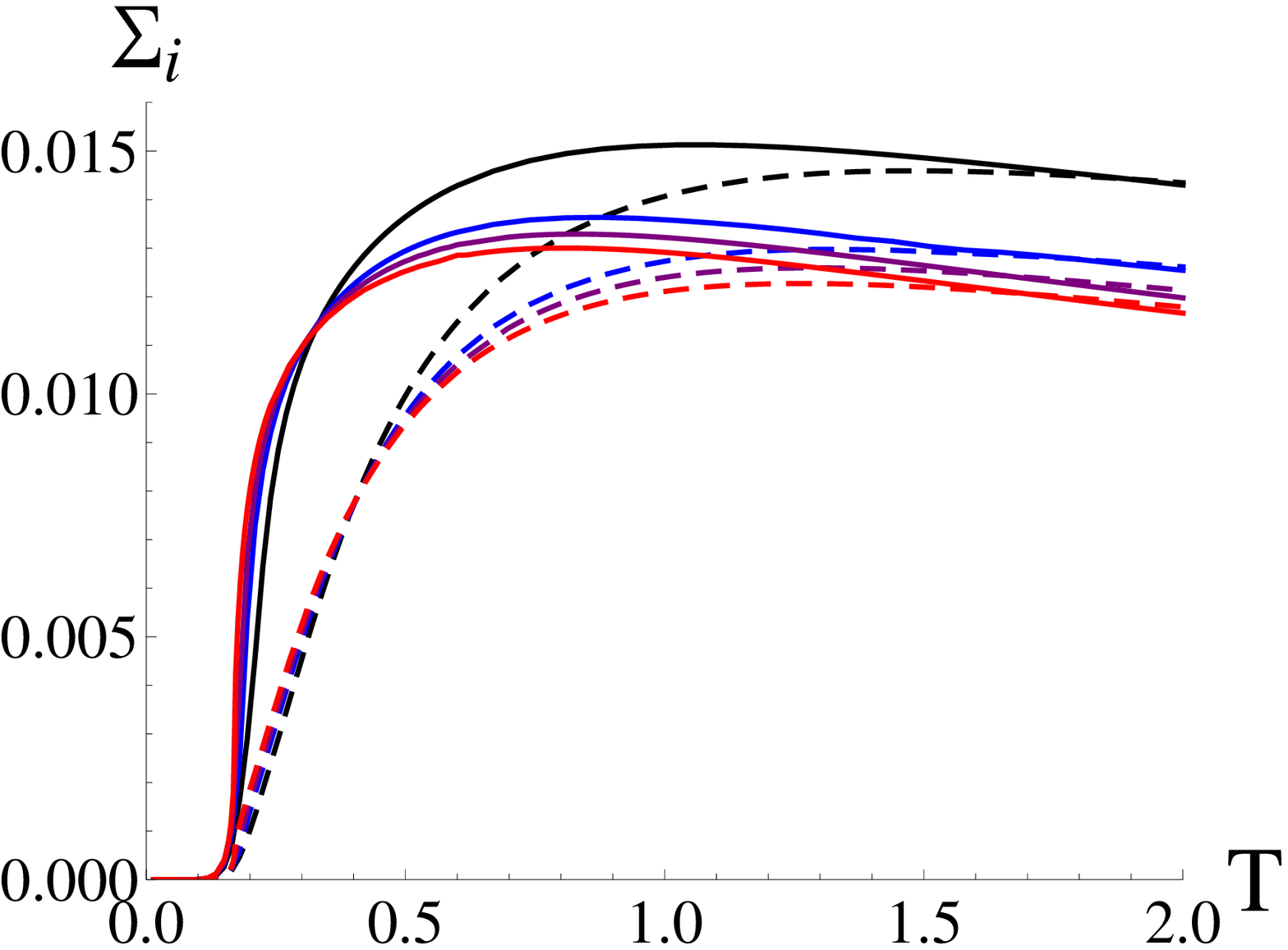}}
\caption{Temperature dependence of: a) the Polyakov loop (full lines starting at zero) and chiral condensates divided by the peak value (light in full lines and strange with dashed lines); b) the dressed Polyakov loop (light in full lines and strange in dashed lines). The different sets correspond to different values for $g_1$ in the PNJL model, $\left.\mathcal{U}\right|_{T_0=190~\mathrm{MeV}}$, $g_1\times 10^{-3}=1, 5, 6.5, 8~\mathrm{GeV} ^{-8}$ (for the chiral condensates and dressed Polyakov loop lower lines correspond to higher $g_1$ whereas for the Polyakov loop it is the opposite).}
\end{figure}
\begin{figure}[htb]
\begin{center}
\subfigure[]{\includegraphics[width= 0.3 \textwidth]{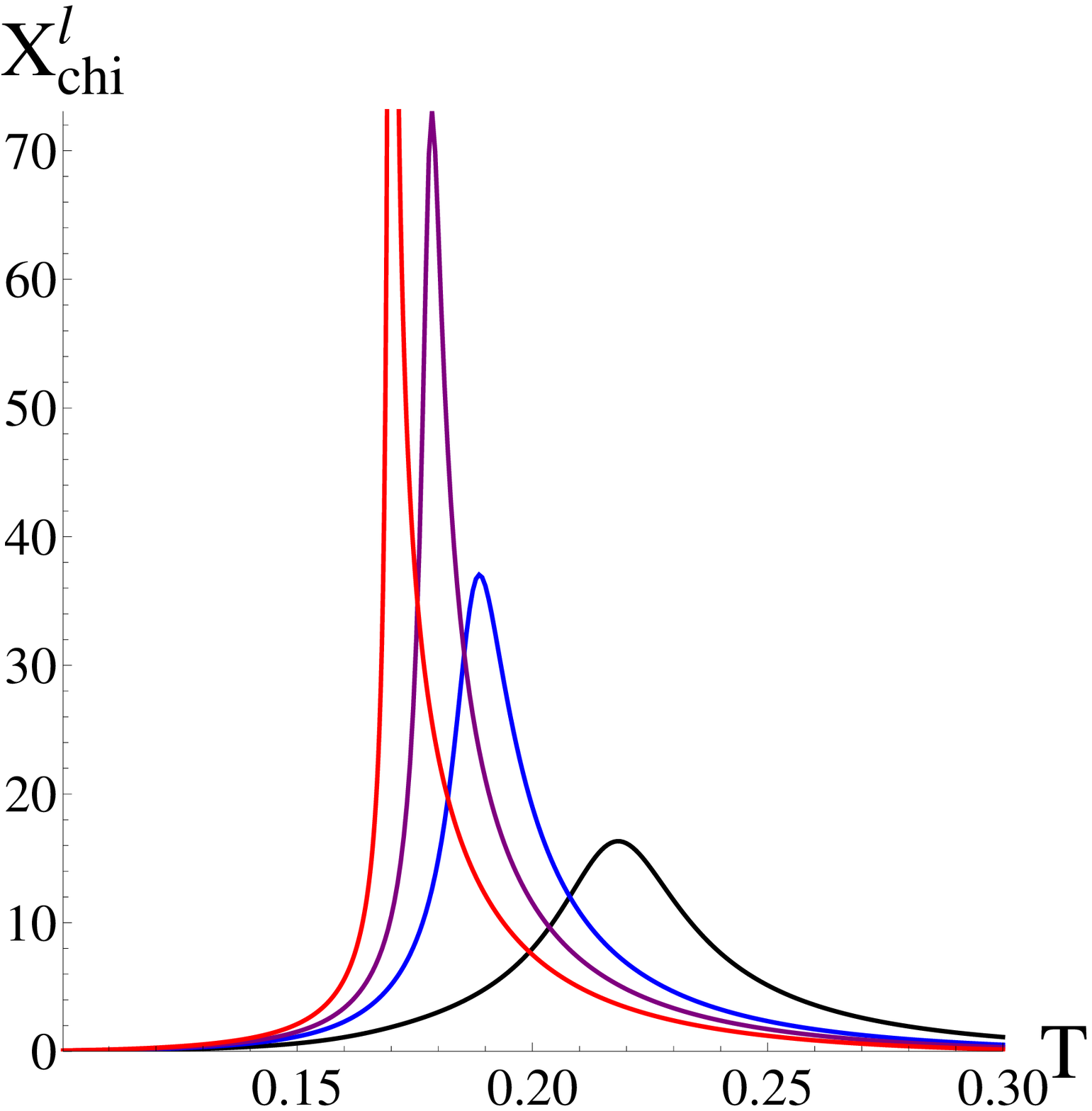}}
\subfigure[]{\includegraphics[width= 0.3 \textwidth]{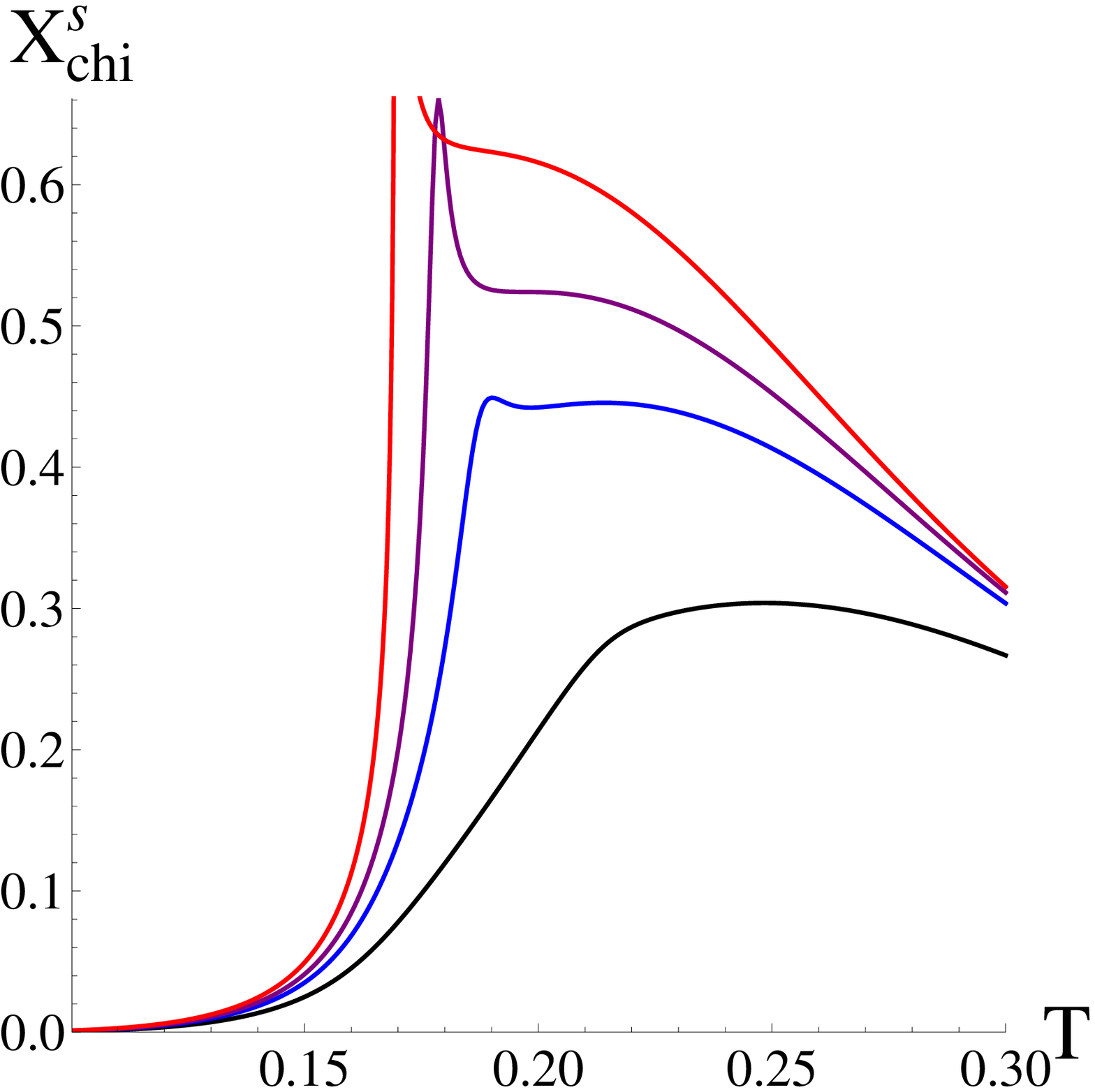}}
\subfigure[]{\includegraphics[width= 0.3 \textwidth]{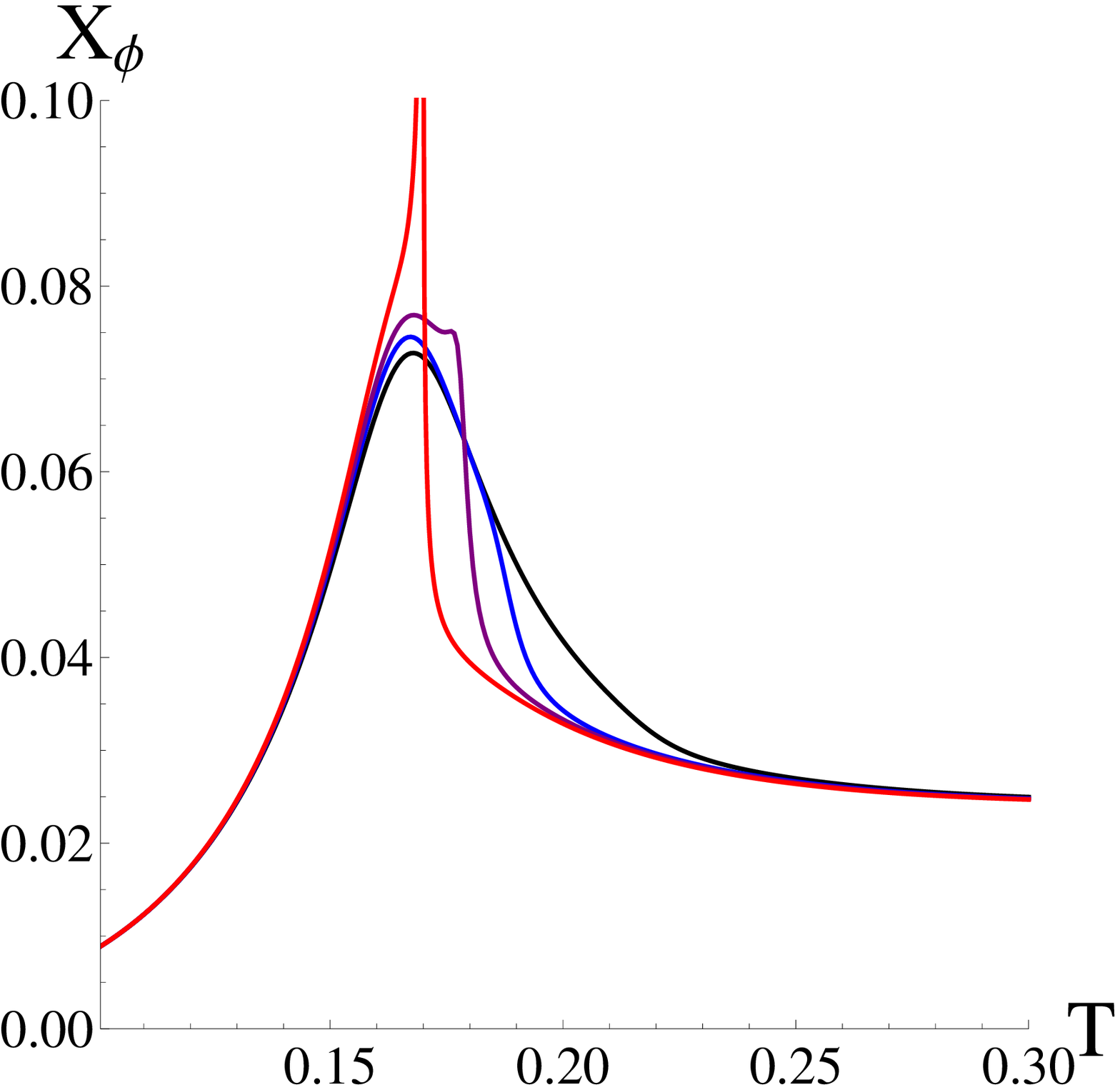}}
\subfigure[]{\includegraphics[width= 0.3 \textwidth]{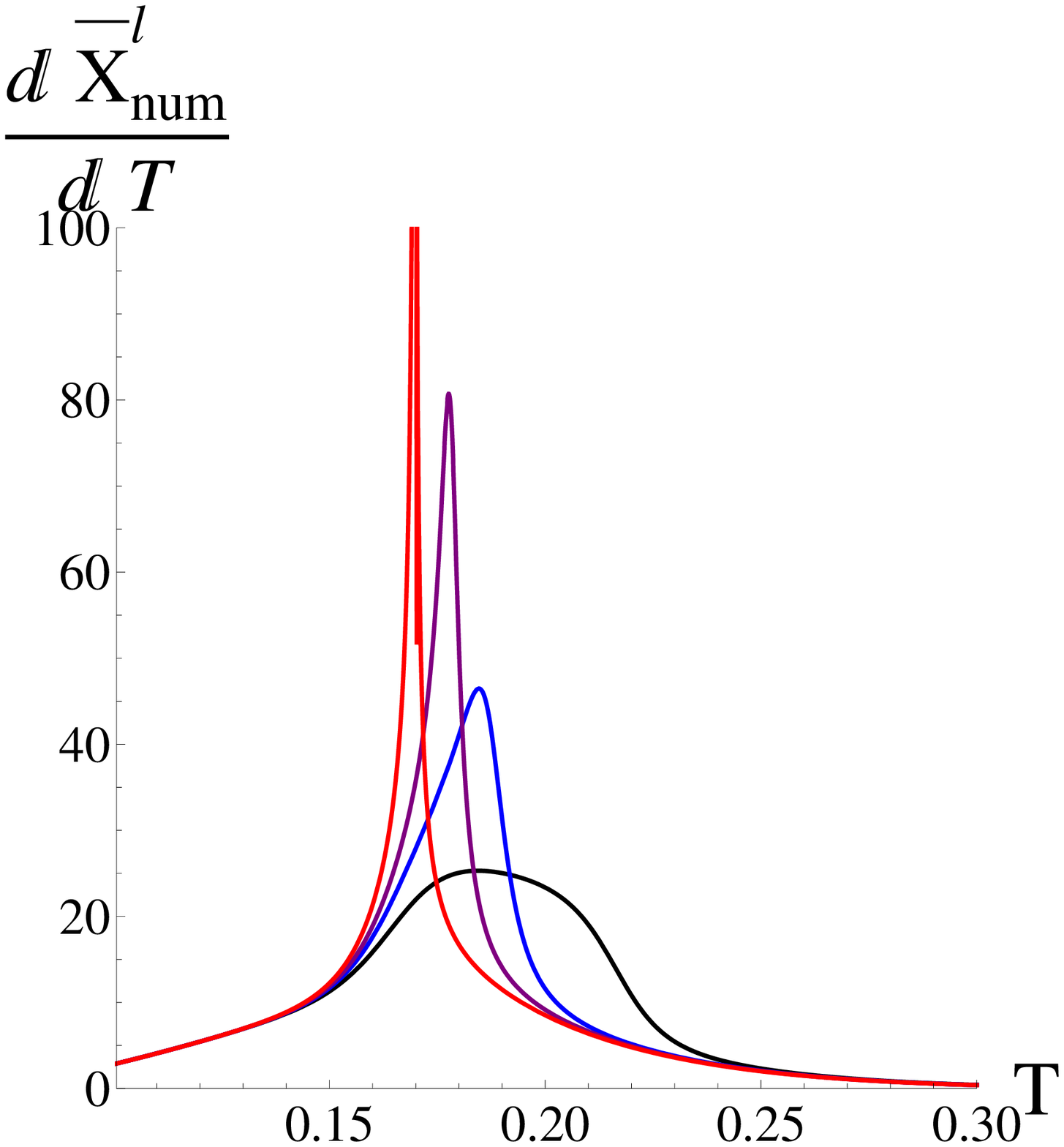}}
\subfigure[]{\includegraphics[width= 0.3 \textwidth]{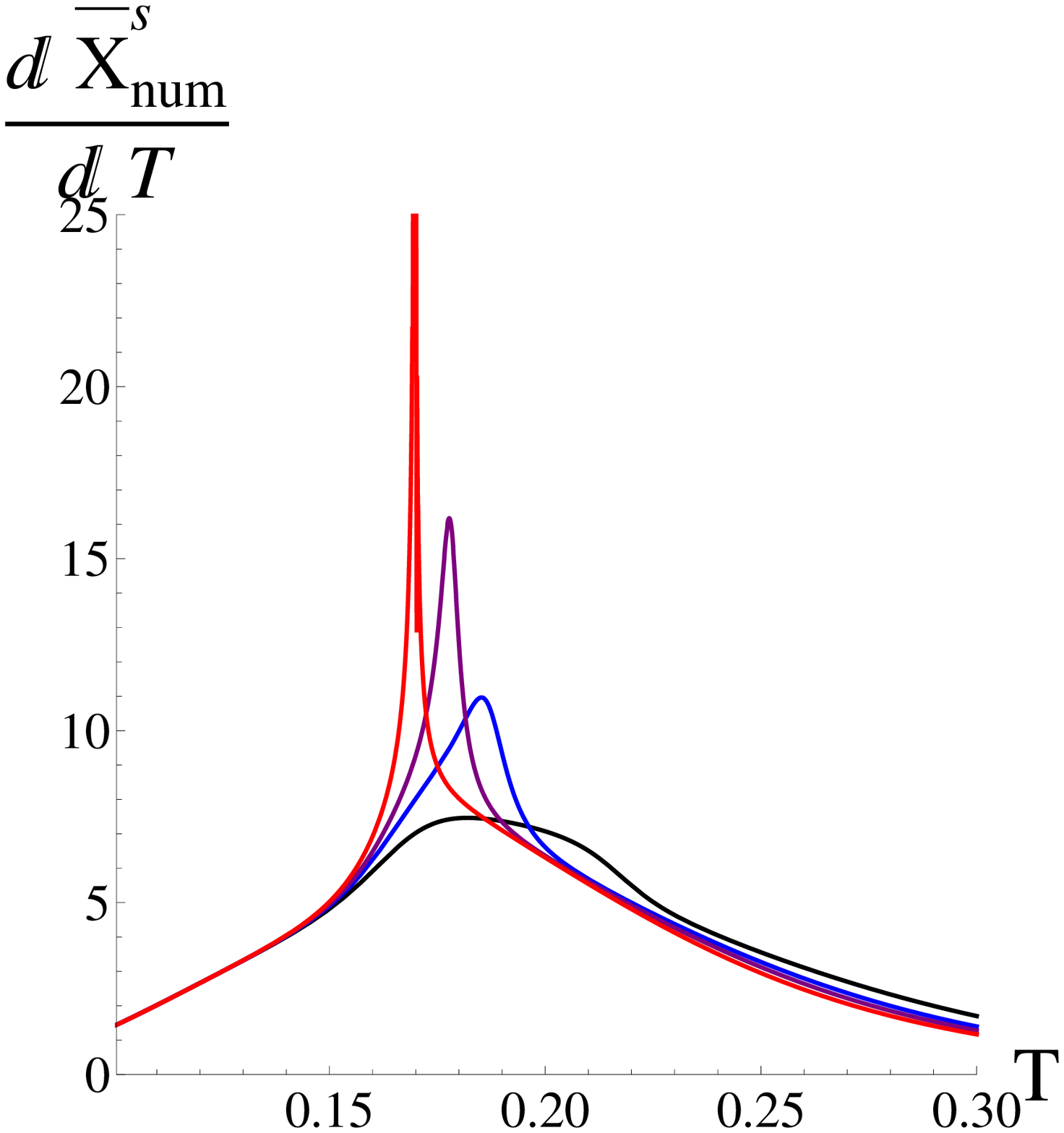}}
\caption{Temperature dependence of susceptibilities in the PNJL model ($\mathcal{U}$ with $T_0=190~\mathrm{MeV}$) for different values of $g_1$ at vanishing chemical potential. Top row from left to right: light quark chiral susceptibility, strange quark chiral susceptibility and Polyakov loop susceptibility. Bottom row: temperature derivative of the light and strange quark number susceptibilities. In all of these panels the sharper peaks correspond to stronger OZI-violating 8q interactions (same sets as in Fig. 1).}.
\end{center}
\end{figure}

A comparison with lQCD results (see Fig. 3 a) and b)) for several quantities such as the interaction measure and the subtracted chiral condensate gives a reasonable result. The deviation from the lQCD result in what refers to the subtracted chiral condensate is probably related to the slow chiral restoration of the strange quark (Fig. 1a). The position of the critical endpoint of QCD in the phase diagram is also very affected by the value of $g_1$ as can be seen in Fig. 3c. The $T_0$ parameter from $\mathcal{U}$ on the other hand only seems to affect the temperature at which it occurs and not the chemical potential. 


\begin{figure}[htb]
\begin{center}
\subfigure[]{\includegraphics[width= 0.28 \textwidth]{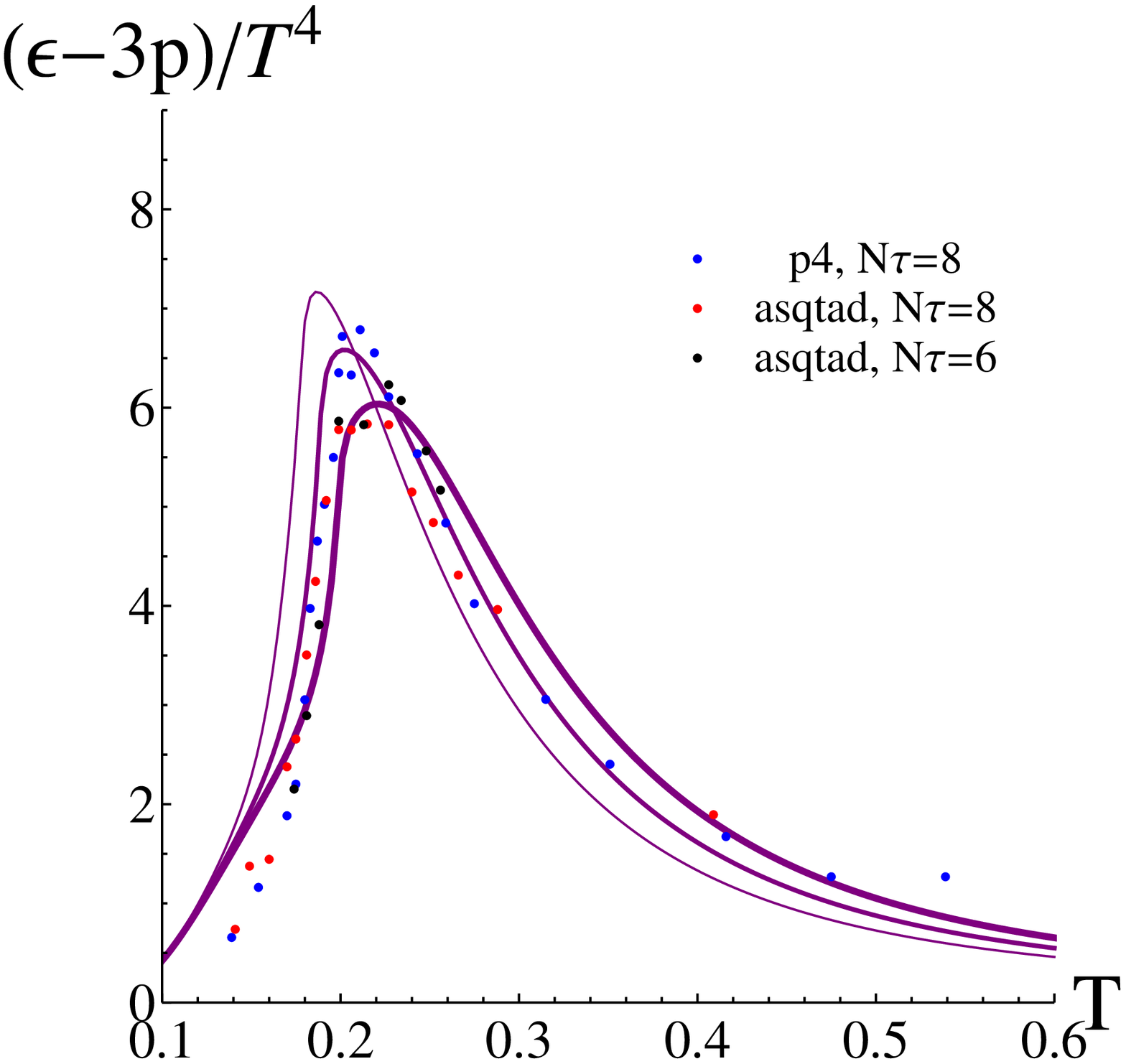}}
\subfigure[]{\includegraphics[width= 0.28 \textwidth]{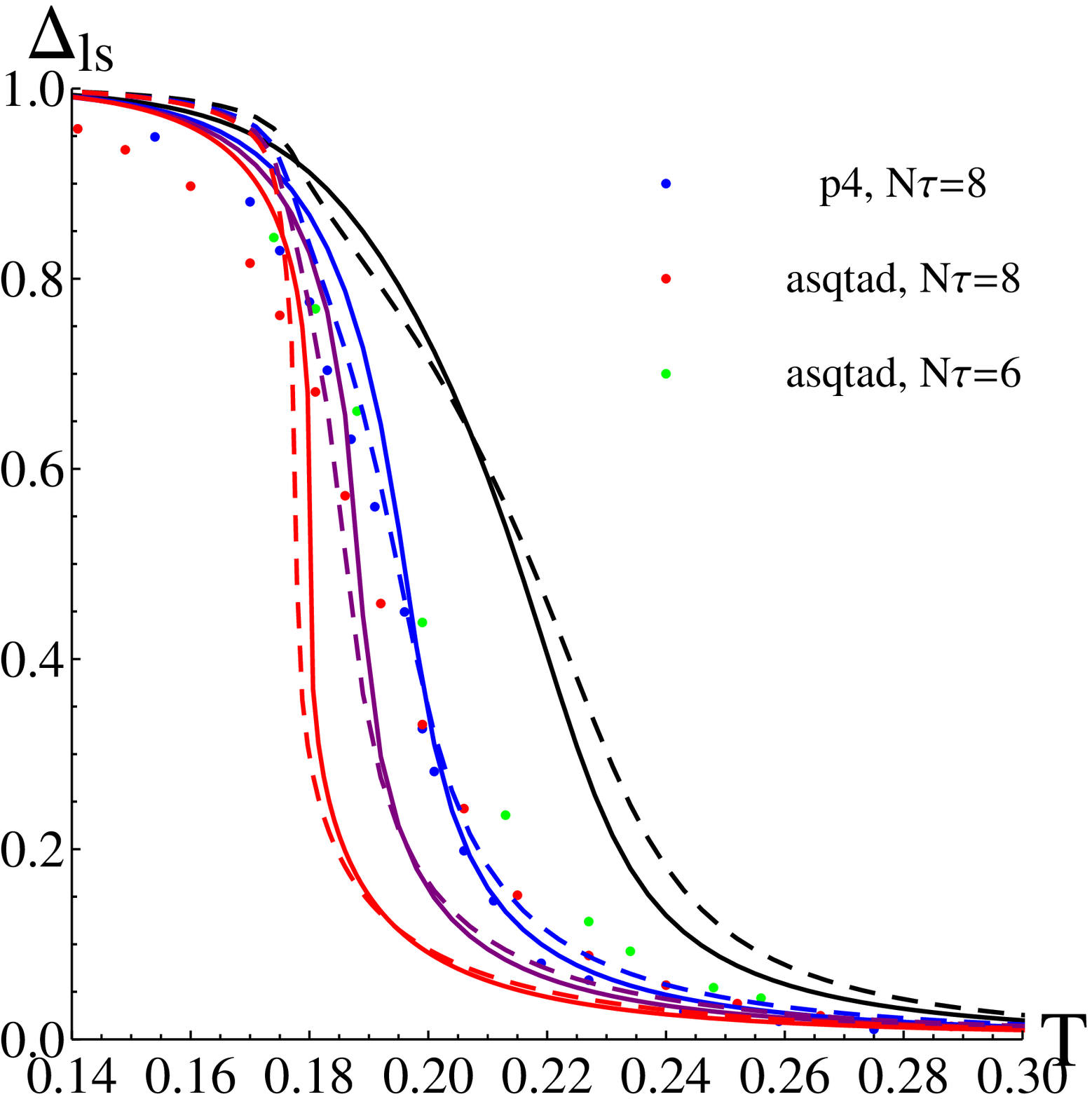}}
\subfigure[]{\includegraphics[width= 0.28 \textwidth]{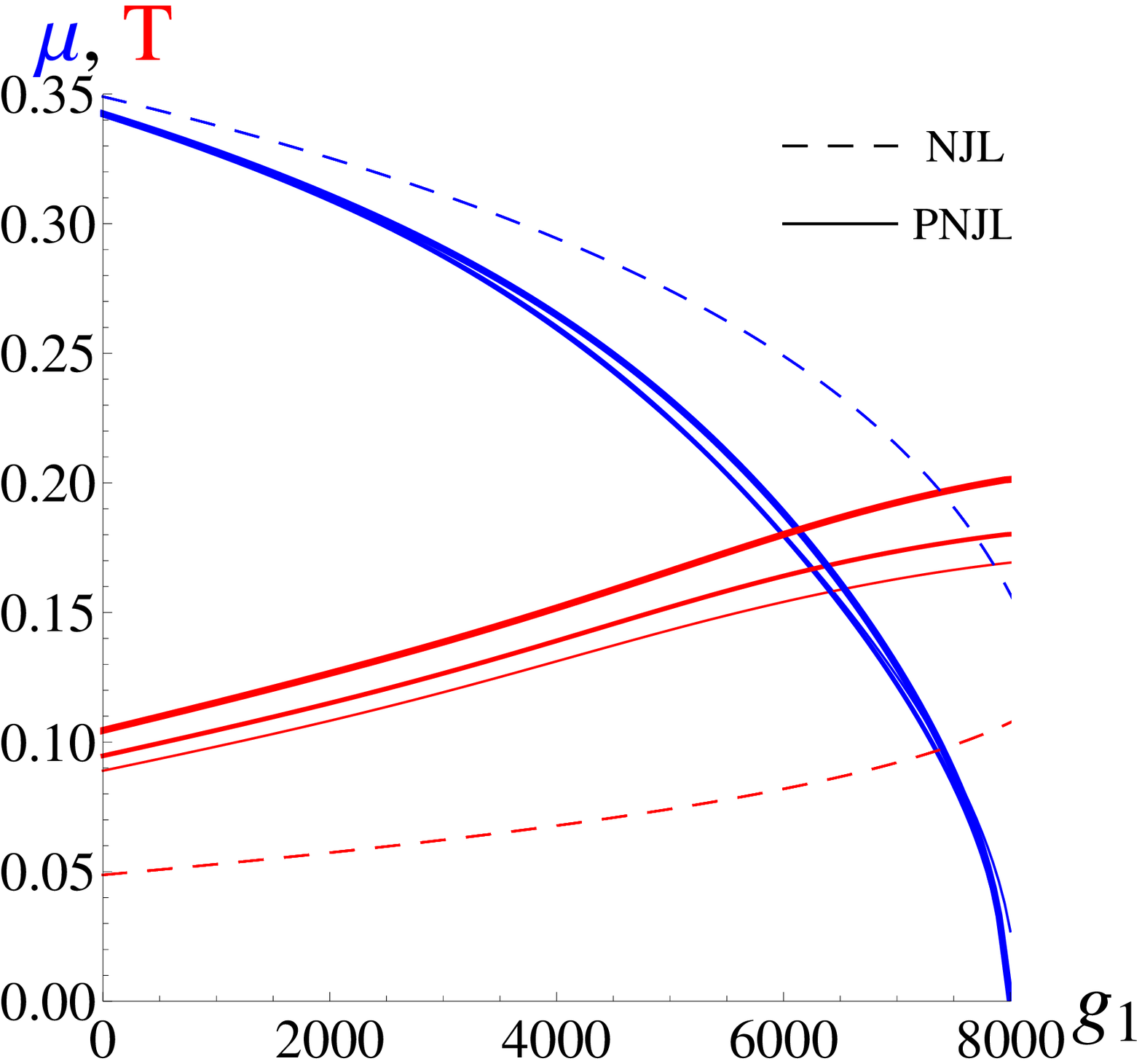}}
\caption{Interaction measure ($g_1=6000~\mathrm{GeV}^{-8}$, $T_0=0.190,~0.210,~0.230~\mathrm{GeV}$; higher $T_0$ results in a lower and more shifted to the right peak) and the subtracted chiral condensate ($T_0=210\mathrm{GeV}$ for comparison we also show in dashed lines the result obtained with the logarithmic form of $\mathcal{U}$ from \cite{Roessner:2007}) compared to lQCD results (data taken from \cite{Bazavov:2009}). Temperature (positive slope lines) and chemical potential (negative slope lines) of the CEP as a function of $g_1$ for $T_0=190,210,250~\mathrm{MeV}$ (higher $T_0$ corresponds to thicker lines) as well as for the NJL model.}
\end{center}
\end{figure}


\section*{Acknowledgements}
 Work supported by FCT, CERN/FP/116334/2010, QREN, UE/FEDER through COMPETE. Part of the EU Research Infrastructure Integrating Activity Study of Strongly Interacting Matter (HadronPhysics3) under the 7th Framework Programme of EU: Grant Agreement No. 283286.

\end{document}